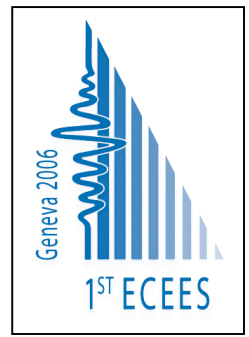

# DYNAMIC BEHAVIOUR OF THE FIRST INSTRUMENTED BUILDING IN FRANCE: THE GRENOBLE CITY HALL

Clotaire MICHEL[1] and Philippe GUEGUEN[2]

## SUMMARY

The French Accelerometric Network (RAP) launched in November 2004 a marked operation for the seismic behaviour assessment of a typical French building. The main goal of this project is to collect accelerometric data in the building and use them to calibrate models or alternative tools used for the seismic behaviour assessment. The final goal of this project is to help the vulnerability assessment of cities in moderate seismic hazard countries. The French Accelerometric Network (RAP) chose to install a permanent network of 6 accelerometers recording continuously the vibrations of the Grenoble City Hall at the basement and at the top. The 13-story building is a RC shear walls building, typical of the RC structures designed at the end of the 60's in France. All the data collected in the building are available on the online access database of the RAP. In addition to the permanent network, an ambient vibration experiment has been performed in 36 points of the whole building. Using the Frequency Domain Decomposition method, these data allowed estimating precisely the different modes of vibration of the structure for low amplitudes. Only the first bending modes in each direction (1.15 and 1.22 Hz) and the first torsion mode (1.44 Hz) are excited. We compared the frequencies obtained using ambient vibration to those for a moderate earthquake recorded by the permanent network. Thanks to the continuous recording, a statistical approach of the torsion mode pointed out the position of the centre of rotation of the building. A modal model extracted from ambient vibrations is proposed and validated thanks to the earthquake recordings collected in the building during the ML=4.6, September 8th 2005 Vallorcine (Haute-Savoie, France) earthquake.

## 1. INTRODUCTION

In moderate seismic countries, the stress has often been put on the hazard estimation so that the vulnerability of the building stock remains unknown. Whereas in California each new high-rise is instrumented with accelerometers, the first building in France has been instrumented in November 2004 with 6 sensors, 3 at the roof and 3 on the ground floor, the network having been planned by the French Accelerometric Network (RAP). This building is the Grenoble City Hall, a 13-story RC-structure built at the end of the 60's as many other Grenoble buildings due to the increase of the population. This structure is therefore typical of a large set of the building stock found in Grenoble and in France. For 1.5 year of continuous recordings, many small alpine earthquakes coming from Italy or Northern Alps have been recorded. The structure has not suffered damages, even before the instrumentation. This building network has been decided in order to get a test building on which experimental and/or numerical approaches might be applied and validated with the accelerometric data.

In addition, we recorded ambient vibrations in the building with a temporary network. All these recordings allow us to evaluate the dynamic behaviour of the structure, with a special focus on the determination of its structural modes. For this purpose, modal analysis have been performed using ambient vibration and moderate earthquake recordings corresponding to the recent ML=4.6 Vallorcine earthquake. In this paper, we pay attention to the estimate of the torsion mode and the corresponding centre of rotation. Finally, we propose a simple modal model based on lumped-mass modelling to describe the motion of each floor, extracted from ambient vibrations and validated with the Vallorcine earthquake.

---

[1] University of Grenoble LGIT Maison des Géosciences, 1381 rue de la Piscine, F – 38041 GRENOBLE, France
Email : clotaire.michel@obs.ujf-grenoble.fr
[2] University of Grenoble LGIT/LCPC Maison des Géosciences, 1381 rue de la Piscine, F – 38041 GRENOBLE, France
Email: philippe.gueguen@obs.ujf-grenoble.fr

## 2. THE GRENOBLE CITY HALL AND THE RECORDING NETWORKS

### 2.1 Presentation of the structure

The Grenoble City Hall is a reinforced-concrete (RC) structure built in 1967 (Fig. 1). It is divided into 2 parts: a 3-level horizontal building and an independent 12-story tower that we will study here. The tower bears on two vertical piles containing the stairs and lifts. The glass frontage is based on a light steel framework. As a consequence, the two piles are controlling the behaviour of this structure and despite its dimensions (44 m length x 13 m wide x 49 m height) the longitudinal and transverse directions should behave almost the same way. The building seems to be symmetric from geometrical and design points of view, so that no torsion should occur in it. For this reason, other specialists [Ma and Mazars, 2004] decided to build a 2D numerical model of this building to explore its non-linear behaviour.

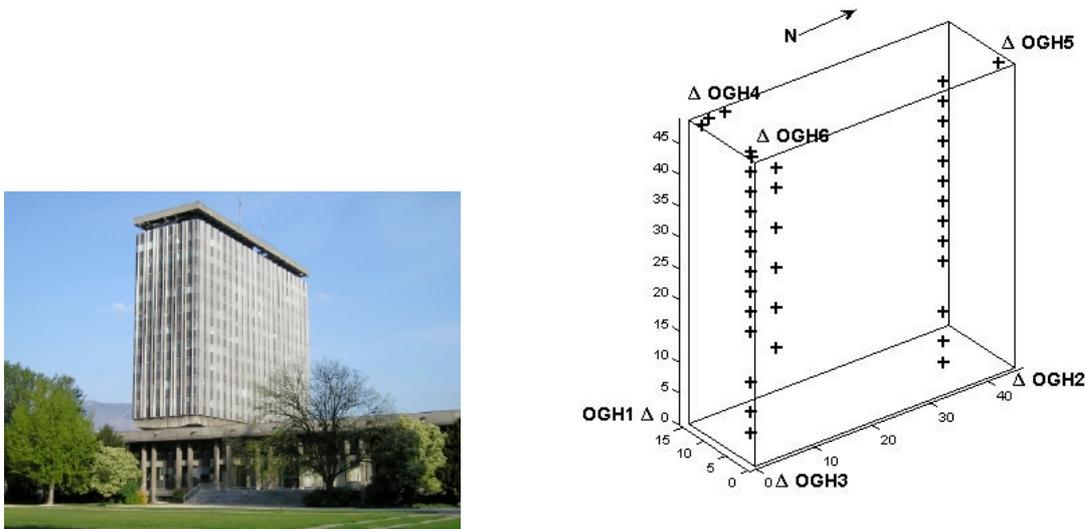

**Figure 1: a) The Grenoble City Hall ; b) The location of the accelerometers part of the RAP (triangles) and the location of the temporary network used for the ambient vibration experiment (plus)**

### 2.2 Permanent network

Since November 2004, 6 accelerometric stations (3 on the basement, 3 at the top) are recording continuously the building vibrations (Fig. 1). This instrumentation is part of the French Accelerometric Network (RAP) policy, which decides marked operations to focus on particular scientific subjects. The scientific board of the RAP decided then to pay attention to building instrumentation in order to contribute to the seismic vulnerability assessment in France, by the experimental way. The City-Hall network is managed by the Geophysical Laboratory of the Grenoble University (LGIT). The sensors are 3C FBA-EST (Kinemetrics) accelerometers, each one connected to a 24 bit digital acquisition system miniTitan (Agecodagis). The sampling rate is 125Hz and the recordings are divided into files of 2 min length. The six stations are independent but connected together by an Ethernet hub that allows the data transfer from each station to the computer centre located at the basement. Besides the control and the management of the acquisition parameters, the computer is online (ADSL line) and the data are then downloaded from the laboratory through *ssh* protocol.
Since 2004, more than 25 events have been recorded within the building. All the data are converted in SAC, ASCII and SEED formats and they are integrated to the online web access database of the RAP (http://www-rap.obs.ujf-grenoble.fr). Only the time windows corresponding to the seismic events detected by the national and regional seismic networks (RéNaSS and Sismalp) are collected that allows the download of the strongest events but also of some ambient vibration windows.

**2.3 Ambient vibration experiment**

In June 2005, a temporary network has been installed to determine the full-scale behaviour of the structure under ambient vibrations. A Cityshark II station [Chatelain et al., 2006] has been used that allows the simultaneously recording of 18 channels. The sensors used were 6 Lennartz 3D 5s velocimeters, having a flat response between 0.2 and 50 Hz. We recorded eight datasets, corresponding to 36 different points of the building that is to say at least 2 points per floor (Fig. 1). One sensor was kept immovable at the building top as reference for all the sets. Because the first frequency had been estimated close to 1 Hz, 15 min of recording time were selected for each set, corresponding to more than 1000 periods, at a 200 Hz sampling rate.

### 3. MODAL ANALYSIS OF THE STRUCTURE

**3.1 Ambient vibration recordings**

In order to extract the modal parameters of the structure from the ambient vibration recordings, we used the Frequency Domain Decomposition (FDD) method [Brincker, 2001]. The idea of this method is to calculate the Fourier Transforms of the correlation matrices for each dataset (Power Spectral Density matrices) and then to perform a singular value decomposition at each frequency. As only 1 or 2 orthogonal structural modes have energy at one particular frequency, the first singular value shows peaks corresponding to the structural modes. Therefore the peak frequencies of the first singular value give the resonance frequencies (Fig. 2) and the first singular vectors give the corresponding modal shapes.

Only 3 modes have been accurately determined (Fig. 2): the first longitudinal mode at 1.15 Hz, the first transverse mode at 1.22 Hz and the first torsion mode at 1.44 Hz. Figure 3 displays interpolated pictures of these shapes. The second modes can hardly be distinguished at 4.5 Hz and 5.7 Hz in the longitudinal and transverse direction, respectively.

As expected, the values of bending frequencies are very close one another. The longitudinal direction is even "softer" than the transverse one. The modal shapes look like bending beams and the frequency ratios $f_2/f_1$ are equal to 3.9 and 4.7 in the longitudinal and transverse direction, respectively. These ratios correspond in the Timoshenko beam model [Boutin et al., 2005, Michel et al. 2006] to non-shear behaviour.

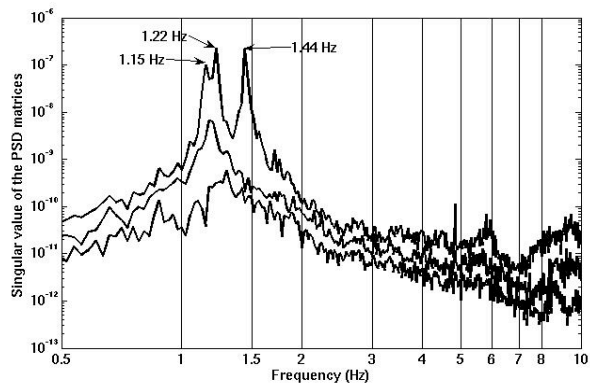

**Figure 2: Spectrum (3 first singular values of the PSD matrices) of the structure under ambient vibrations**

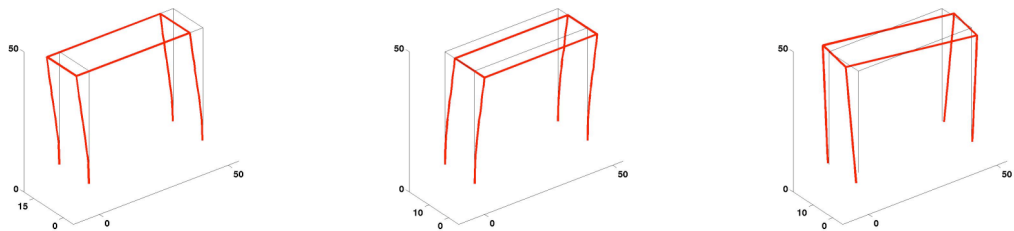

**Figure 3: 3 first structural modes of the structure (from left to right: longitudinal bending, transverse bending and torsion)**

## 3.2 Earthquake recordings

Thanks to the permanent network, the ML=4.6, September 8th 2005 Vallorcine (Haute-Savoie, France) earthquake has been recorded. Only minor damage and rock falls occurred in the Chamonix Valley but it was strongly felt in the Alps and especially in Grenoble basin, certainly due to the strong site effects [Lebrun et al., 2001]. This event is the strongest event recorded since the permanent recording started in the Grenoble City Hall. Even if no damage was observed in Grenoble, more than 100 km from the epicentre, people working beyond the third level spontaneously evacuated the City Hall.

Although the basic assumption of white noise is required for the FDD method, it was also used to determine the modes of the structure. The FDD is robust enough to allow this process [Ventura et al., 2003]. We found a slight decrease in the first frequencies from 2% to 4% (Tab. 1). This decrease may be due to the aperture of micro-cracks in the concrete that slightly decrease the stiffness of the structure and therefore the frequencies, as already mentioned by Dunand [2004] using Californian data collected in buildings.

**Table 1: Comparison between the resonance frequencies of the structure under ambient vibrations and under Vallorcine earthquake**

| Resonance frequencies | Ambient vibrations | Vallorcine earthquake | Decrease |
|---|---|---|---|
| $1^{st}$ longitudinal (Hz) | 1.15 | 1.13 | 2% |
| $1^{st}$ transverse (Hz) | 1.22 | 1.17 | 4% |
| $1^{st}$ torsion (Hz) | 1.44 | 1.42 | 1.5% |

## 3.3 Centre of rotation

The torsion mode determined using ambient or earthquake vibration is quite pure so that it is not coupled with any translation motion. Assuming this mode is a rigid body rotation, it is then possible to determine its centre. As we have 3 points on the roof with their 3D directions, this problem is overdetermined. The best dataset we can use for these computations comes from the permanent network because the RAP stored many recordings of ambient vibrations thanks to the continuously recording mode. For each recording, we applied the FDD and selected the torsion mode. We then calculated the coordinates of the best centre C(x,y) assuming a rigid body rotation, after the method detailed in [Brownjohn, 1996] (Fig. 4).

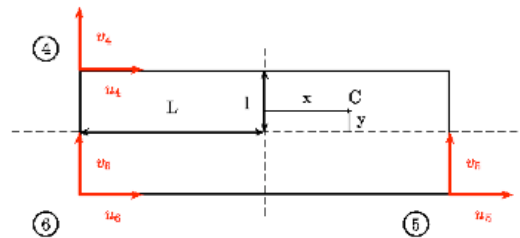

**Figure 4: Schematic planar view of the roof of the Grenoble Town Hall with the permanent stations OGH4, OGH5 and OGH6**

The best synchronized data were used that correspond to about 178 recordings. Statistics on the results give us an idea of the reliability of this method (Tab. 2, Fig. 5).

**Table 2: Statistics of the centre of rotation determination and the torsion frequency**

|  | X coordinate (m) | Y coordinate (m) | Frequency (Hz) |
|---|---|---|---|
| Mean | -0.63 | 1.50 | 1.453 |
| Standard deviation | 1.00 | 0.49 | 0.008 |
| Mean uncertainty | 0.15 | 0.07 |  |

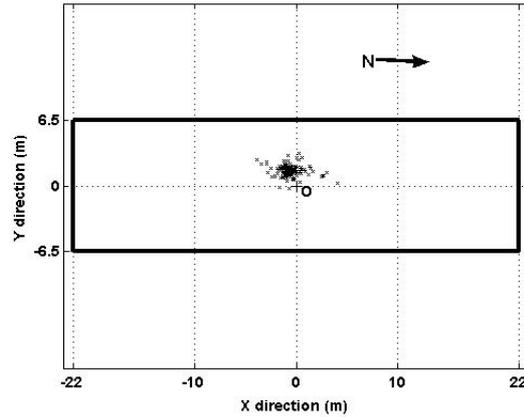

**Figure 5: Spatial distribution of the rotation centres of the building for the 178 recordings**

The frequency standard deviation is depending on the length of the Fourier Transform windows (125 Hz / 16384 points = 0.0076 Hz). The frequency value found is consistent with the previous modal analysis results. The statistics on the centre of rotation coordinates show that this centre has a natural dispersion around a well determined point (-0.63±0.15 m ; 1.50±0.07 m). Looking at the building plans, there is no evidence that the structure may not be symmetric. The ambient vibration test may be therefore useful for the estimate and the detection of the torsion mode.

## 4. SIMPLE MODAL MODELLING

### 4.1 Presentation of the model

In a building, the masses are concentrated at the floors. Therefore we assumed a lumped-mass modelling for this structure. In this case, the Duhamel integral [Clough and Penzien, 1993] gives us the motion of the structure at each floor knowing only the mass of the stories, the vibration modes and the motion of the ground. We assume here a mass of 1000 kg/m$^2$ for each floor (standard values for such a building) and we consider that only the first bending modes provided energy, neglecting the torsion mode for the sake of simplicity. The damping ratio have been set to 1% and 2 % for the longitudinal and transverse direction, respectively, thanks to a random decrement evaluation on ambient vibrations [Brincker, 1991]. It is then possible to compute the motion at each floor for any deterministic earthquake scenario. This is of course a linear model, which suits only for moderate motions. Nevertheless, as mentioned in Boutin et al. [2005], elastic modelling can be used to detect whether the building reaches the post-elastic state or not.

### 4.2 Validation using the ML=4.6, September 8th 2005 Vallorcine (Haute-Savoie, France) earthquake

The Vallorcine recordings (Fig. 6) show a strong anisotropy in the building: the amplitude of the transverse direction (East) is two times the amplitude of the longitudinal direction at the top, which is not usual for the recordings in the City Hall. Even if usually there are 4 independent motions in a structure [Guéguen, 2000] (relative motion of the foundation, base rocking, torsion and bending), for this earthquake we determine that the motion is mostly bending, which confirms the interest of the modal model we propose above. Considering the recording at the basement level as input, we are able to compare at the top level the modelled motion with the recorded one (Fig. 7). The usual parameters describing the motion are well fitted (acceleration amplitude, duration…) by the computed motion. That validates this simple modal modelling extracted from ambient vibrations. Thanks to the lumped-mass model, we show also that the strain at each story (Fig. 8) is greater in the transverse than in the longitudinal direction, especially for the last floor. This strain is quite far (one order of magnitude) from the minimum strain able to damage reinforced concrete [Boutin et al., 2005].

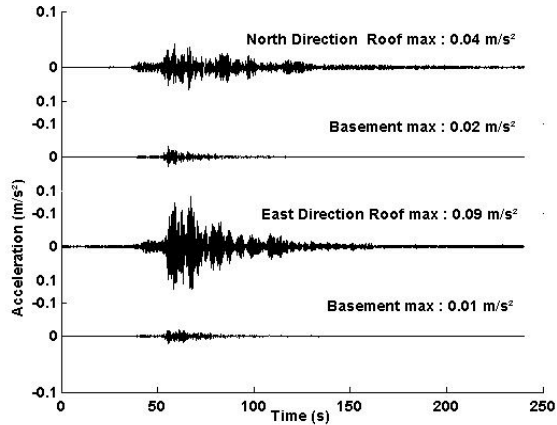

**Figure 6: Vallorcine earthquake recorded in the Grenoble City Hall by the RAP network.**

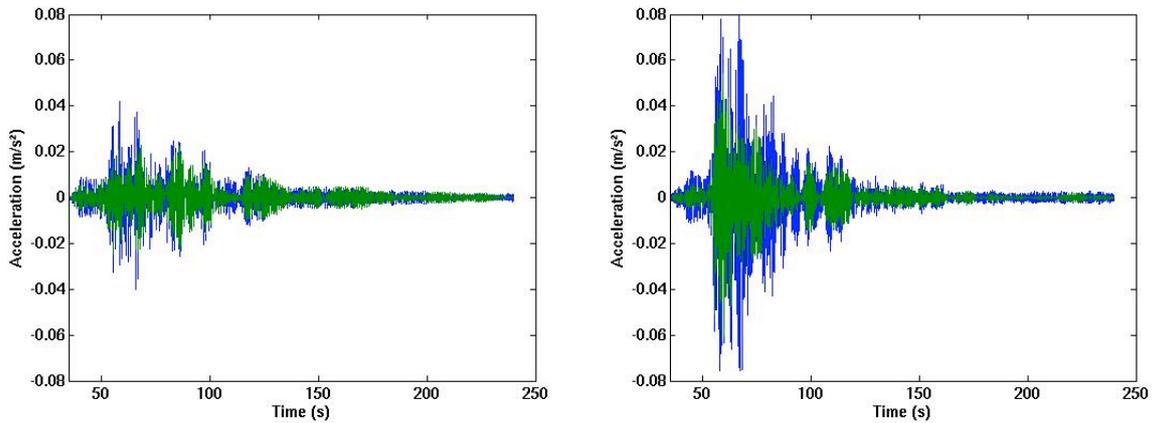

**Figure 7: Comparison between recordings (blue) and modelling (green) of the Vallorcine earthquake at the top of the struture in the North (left) and East (right) directions**

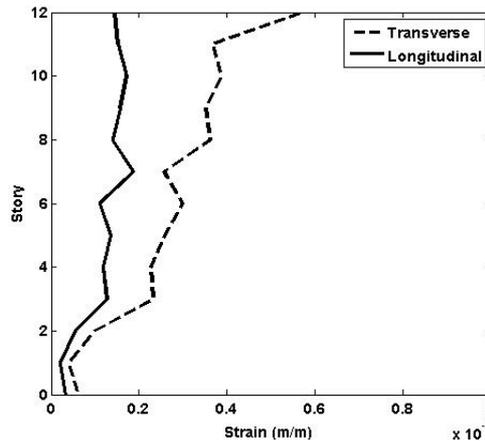

**Figure 8: Modelling of the maximum strain along the stories of the structure during Vallorcine earthquake**

## 5. CONCLUSION

In this paper we showed how modal analysis using ambient vibrations could help understanding the dynamic behaviour of a building such as the Grenoble City Hall. We determined its main vibration modes (bending and torsion) using both earthquake and ambient vibration recordings. We showed that the method used, the FDD method, gave consistent results in both cases. We also computed the centre of rotation of the structure related to a pure torsion mode. It is shifted to the West without apparent reasons on the building plans. We proposed a simple elastic modal model based on ambient vibration modes to describe the motion of each story under earthquakes and validated it thanks to the Vallorcine earthquake recorded by the permanent network (RAP). With this approach, we showed that ambient vibration tests in building can be used for the modal model estimation and then for the modelling of the building motion in case of earthquakes scenarios. This approach allows then the update of the elastic domain of the building model and the estimate of the deformation rate that can be compared to the integrity threshold concept of buildings [Boutin et al., 2005].

## 6. Acknowledgment

This study, part of the VULNERALP project, is supported by the Regional Council of Rhône-Alpes, within the framework of its "Thématiques Prioritaires" scientific program. The accelerometric data are provided by the French Accelerometric Network database (http://www-rap.obs.ujf-grenoble.fr).